\documentclass[letterpaper]{article} 
\usepackage{aaai23}  
\usepackage{times}  
\usepackage{helvet}  
\usepackage{courier}  
\usepackage[hyphens]{url}  
\usepackage{graphicx} 
\def\UrlFont{\rm}  
\usepackage{natbib}  
\usepackage{caption} 
\usepackage{algorithm}
\usepackage{algorithmic}

\usepackage{newfloat}
\usepackage{listings}
\DeclareCaptionStyle{ruled}{labelfont=normalfont,labelsep=colon,strut=off} 
\floatstyle{ruled}
\newfloat{listing}{tb}{lst}{}
\floatname{listing}{Listing}
\newcommand{\DatasetName}{\textsc{InfraCorp}}
\newcommand{\DatasetNameVZero}{\textsc{WHS-Corp}}
\newcommand{\DatasetNameVOne}{\textsc{InfraCorp-A}}
\newcommand{\DatasetNameVTwo}{\textsc{InfraCorp-R}}

\newcommand{\ToolkitName}{\textsc{\textbf{NewsPanda}}}
\newcommand{\ModelName}{\textsc{\textbf{NewsPanda}}}

\newcommand{\BotName}{\textsc{WildlifeNewsIndia}}

\usepackage{xcolor}
\usepackage{amsfonts}
\usepackage{appendix}
\usepackage{caption}
\usepackage{subcaption}

\usepackage[strict]{changepage}

\usepackage{framed}

\definecolor{formalshade}{rgb}{0.95,0.95,1}

\newenvironment{formal}{%
  \MakeFramed{\advance\hsize-\width\FrameRestore}%
  \noindent\hspace{-4.55pt}
  \begin{adjustwidth}{}{7pt}%
  \vspace{0.5pt}\vspace{0.5pt}%
}
{%
  \vspace{0.5pt}\end{adjustwidth}\endMakeFramed%
}

\title{NewsPanda: Media Monitoring for Timely Conservation Action}
\author {
Sedrick Scott Keh\equalcontrib\textsuperscript{\rm 1},
Zheyuan Ryan Shi\equalcontrib\textsuperscript{\rm 1, \rm 4},
David J. Patterson\textsuperscript{\rm 2},
Nirmal Bhagabati\textsuperscript{\rm 3}\thanks{Nirmal Bhagabati was not at USAID when the research for this paper was conducted. The views and opinions expressed in this paper are those of the authors and not necessarily those of USAID.
},\\
Karun Dewan\textsuperscript{\rm 2},
Areendran Gopala\textsuperscript{\rm 2},
Pablo Izquierdo\textsuperscript{\rm 2},
Debojyoti Mallick\textsuperscript{\rm 2},
Ambika Sharma\textsuperscript{\rm 2},
Pooja Shrestha\textsuperscript{\rm 2},
Fei Fang\textsuperscript{\rm 1}
}
\affiliations {
    \textsuperscript{\rm 1}Carnegie Mellon University, \\
    \textsuperscript{\rm 2}World Wide Fund for Nature, \\
    \textsuperscript{\rm 3}United States Agency for International Development, \\
    \textsuperscript{\rm 4}98Connect 
}

\usepackage{bibentry}

\begin{document}

\maketitle

\begin{abstract}
Non-governmental organizations for environmental conservation have a significant interest in monitoring conservation-related media and getting timely updates about infrastructure construction projects as they may cause massive impact to key conservation areas. Such monitoring, however, is difficult and time-consuming. We introduce {\ToolkitName}, a toolkit which automatically detects and analyzes online articles related to environmental conservation and infrastructure construction. We fine-tune a BERT-based model using active learning methods and noise correction algorithms to identify articles that are relevant to conservation and infrastructure construction. For the identified articles, we perform further analysis, extracting keywords and finding potentially related sources. {\ToolkitName} has been successfully deployed by the World Wide Fund for Nature teams in the UK, India, and Nepal since February 2022. It currently monitors over 80,000 websites and 1,074 conservation sites across India and Nepal, saving more than 30 hours of human efforts weekly. We have now scaled it up to cover 60,000 conservation sites globally. 
\end{abstract}

\section{Introduction}

Massive floods, poaching, waste pollution -- every week, new threats impacting our environment come to light. Each of these can cause a long chain of negative impacts if not addressed. As such, monitoring these conservation-related events is of great importance for non-governmental organizations (NGOs) focused on environmental conservation such as the World Wide Fund for Nature (WWF) to take timely action and participate in relevant conversations. 

In addition to conservation as a whole, many NGOs are particularly interested in monitoring news on certain subtopics. One such area is the ongoing or upcoming infrastructure projects such as roads, railways, and pipelines. These are usually more long-term and actionable than events like disasters or animal activity which occur in the past or present (hence limiting intervention impact).
Conservation NGOs such as WWF play a key role in advocating for more sustainable infrastructure development.
Early detection and engagement of these projects could shift infrastructure planning towards more environmentally sustainable outcomes while benefiting the people that the projects intend to serve. 

However, information about conservation-related events and infrastructure plans threatening critical habitats is scattered across numerous sources and comes in different forms. 
NGOs typically learn of such information through word-of-mouth or a handful of news outlets that they check manually. This process is both time-consuming and ineffective, and it can potentially fail to capture critical information in a timely manner, leaving these NGOs out of key conversations during early or ongoing stages of these developments. 

\begin{figure}
    \centering
    \includegraphics[width=0.47\textwidth]{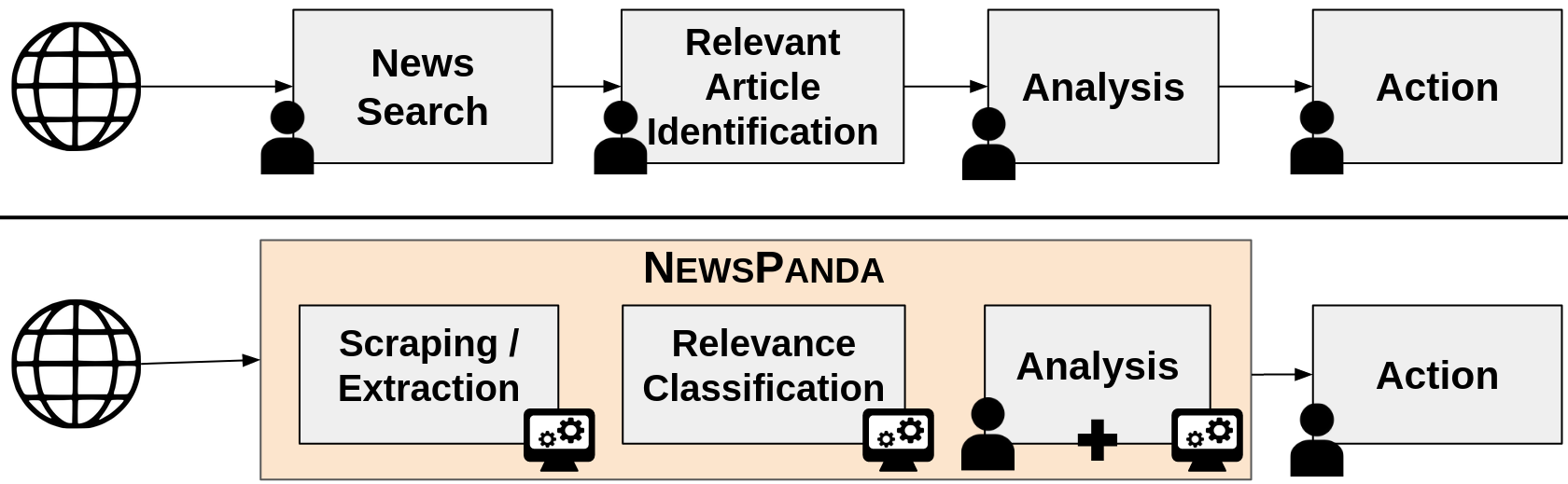}
    \caption{Top: Current costly and time-consuming information gathering pipeline at NGOs. Bottom: \textsc{NewsPanda} automates multiple steps in the pipeline, enabling humans to perform the more critical tasks (analysis and action). }
    \label{fig:page-1-fig}
\end{figure}

To fill this gap, we develop {\ToolkitName}, a natural language processing (NLP) toolkit to automatically detect and analyze news and government articles describing threats to conservation areas. {\ToolkitName} has five main components, which we detail in Section \ref{sec:overview}. At the core of {\ToolkitName} is a classification module built using a BERT-based language model, which we fine-tune to classify whether articles are relevant to conservation and to infrastructure.

Developing such a tool in the conservation nonprofit setting poses several unique challenges. First, labeling data is expensive. We propose an active learning-based method to selectively acquire labels on the most critical data points. Second, the data labels could be noisy since labeling for relevance is ultimately a subjective judgement, even if we fix a labeling rubric. We adopt a noise reduction algorithm \cite{DBLP:conf/iclr/ChengZLGSL21} to improve our model's performance.

{\ToolkitName} was developed as a collaboration between WWF and Carnegie Mellon University (CMU). It has been successfully deployed since February 2022 and has been used by the WWF teams in the UK, India, and Nepal to monitor developments in conservation sites. The entire pipeline runs on a weekly basis, scraping and classifying relevant news articles regarding conservation and infrastructure construction related events that occurred in the past week. These articles are then visualized in WWF's GIS systems for the field teams to investigate. We also share some results through social media for the benefit of the broader civil society. Through the deployment of {\ModelName}, the WWF teams have been able to save over 30 hours weekly on collecting news, which allows us at WWF to instead focus on analyzing the news and taking actions (Figure~\ref{fig:page-1-fig}) 
\footnote{We are happy to work with interested researchers and nonprofits on sharing our code and data.}.
\section{Related Work}

\subsubsection{News Monitoring Systems}
\label{sec:rrl-news-monitoring}
Although there is a rich literature on news information extraction in general domains~\cite{Ojokoh2012, 10.1145/988672.988740} as well as some specific applications~\cite{omdena,Joshi2016}, there has been hardly any media monitoring tool for environmental conservation and infrastructure construction. 
Directly using generic media monitoring tools often lead to unsatisfactory results that are not localized enough to be actionable for a specific conservation site or not relevant enough to be reliable.
As a result, conservation NGOs still use a manual process to collect articles.
The only work on conservation news monitoring that we are aware of is a preliminary attempt by~\citet{data_study_group_team_2020_3878457} that apply BERT to classify news articles. Compared to that, with \ToolkitName{} we provide a classification module with algorithmic contributions to address challenges in using the tool in the nonprofit context, a full end-to-end information extraction and processing pipeline, and most importantly, results and lessons learned from a large scale deployment of the tool.
This is the first comprehensive and actionable media monitoring tool for conservation and infrastructure.

\subsubsection{NLP for Conservation \& Infrastructure}
\label{sec:rrl-nlp}
Outside of news monitoring, NLP tools have been used for various applications in conservation and infrastructure. Some analyze the relevant news articles for general insights on conservation reporting~\cite{10.1093/biosci/biaa175} or study their spread and impact~\cite{wu2018using}. These studies are descriptive in nature and orthogonal to our work.
The few studies that take the civil society stakeholder's perspective are focused on different links in the process from us. \citet{luccioni2020analyzing} use BERT-based models to analyze corporate environment sustainability reports.  \citet{Boutilier2020} explore mining-related texts to analyze the social license of a particular project. They target different problems from us. They assume a relevant text is readily available and try to extract meaningful insights from it. On the other hand, we work on identifying that relevant text from thousands of irrelevant texts in the first place and leave the insight extraction to professional organizations like WWF that have been doing that for years.
\begin{figure*}
    \centering
    \begin{subfigure}[b]{0.7\textwidth}
         \centering
         \includegraphics[width=\textwidth]{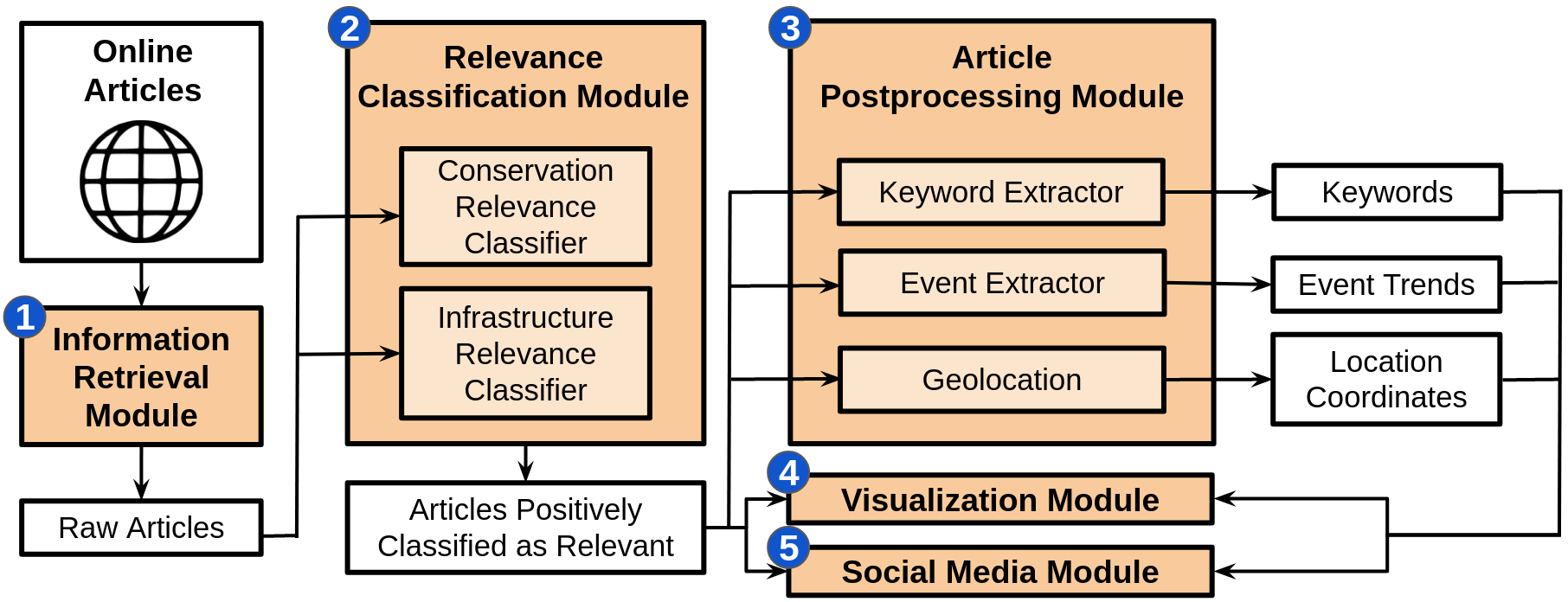}
         \caption{Diagram of overall {\ToolkitName} pipeline, with the five key modules in orange boxes. Generated outputs of {\ToolkitName} are in the white boxes.}
         \label{fig:pipeline-fig}
     \end{subfigure}
     \hfill
     \begin{subfigure}[b]{0.25\textwidth}
         \centering
         \includegraphics[width=\textwidth]{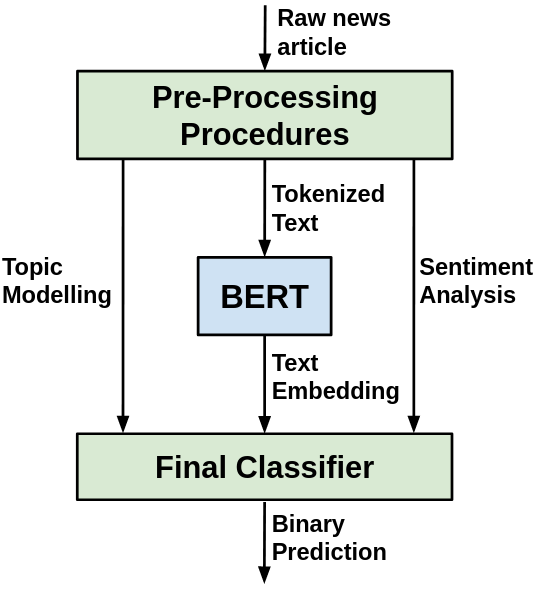}
         \caption{Conservation and infrastructure classification models.}
         \label{fig:model-fig}
     \end{subfigure}
     \hfill
    \caption{\textsc{NewsPanda} pipeline (\ref{fig:pipeline-fig}) and model diagram for conservation and infrastructure relevance classifiers (\ref{fig:model-fig}).}
    \label{fig:main-pipeline-model}
\end{figure*}

\section{{\ToolkitName} Overview}
\label{sec:overview}
{\ToolkitName} toolkit consists of five modules as illustrated below and in Figure \ref{fig:pipeline-fig}. During pilot study and deployment (Section \ref{sec:deployment}), this entire pipeline is run on a weekly basis.
\begin{enumerate}
    \item \textbf{Information Retrieval Module}: We use the \texttt{NewsAPI} scraper~\cite{newsapi} with the names of conservation sites taken from a curated list of conservation areas.
    \item \textbf{Relevance Classification Module}: We classify articles along two dimensions, namely \textit{Conservation Relevance} and \textit{Infrastructure Relevance}, through a large pretrained language model fine-tuned with our collected dataset. Details of this model are explained in Section \ref{sec:model}.
    \item \textbf{Article Postprocessing Module}: The article postprocessing module has 3 parts: a keyword extractor which extracts keywords, an event extractor which extracts event trends, and a geolocator which provides location coordinates. We discuss these features in Section~\ref{sec:article-postprocessing-module}.
    \item \textbf{Visualization Module}: After the relevant articles are identified, we visualize them in our GIS system at WWF, which we can further analyze and act upon  (Section \ref{sec:deployment}).
    \item \textbf{Social Media Module}: In parallel to the visualization module, another downstream application for {\ModelName} is {\BotName}, \footnote{\texttt{https://twitter.com/WildlifeNewsIND}} a Twitter bot we built from {\ModelName} that shares weekly relevant conservation-related articles on social media (Section \ref{sec:deployment}).
\end{enumerate}
\section{Dataset}
\label{sec:dataset}
We use two main datasets for developing {\ModelName}. First, we use an existing corpus ({\DatasetNameVZero}) by \citet{data_study_group_team_2020_3878457} consisting of articles scraped using World Heritage Sites as keywords and labelled by domain experts. Second, we scrape and label our own corpus ({\DatasetName}), which is a more focused, timely, and fine-grained upgrade over {\DatasetNameVZero}. The datasets differ in terms of the locations of the conservation sites used, as well as the time frame of the articles.

\subsection{{\DatasetNameVZero} Dataset}
\label{sec:datasetv0}
{\DatasetNameVZero} contains over 44,000 articles from 2,974 different sources covering 224 World Heritage Sites globally. Scraping was done using \texttt{NewsAPI}'s Python library from a list of curated conservation sites of interest. 
Besides the title and content, it also contains metadata such as the publication site, the author, and the date of publication. Articles in {\DatasetNameVZero} span from January 2018 to October 2019. 

After these articles were gathered, a subset of 928 articles were sampled and manually annotated for \textit{Conservation Relevance} by domain experts familiar with conservation. \textit{Conservation Relevance} denotes whether an article discusses threats or impacts to wildlife and environment conservation in general, e.g. poaching, forest development, natural disasters. 
We use this labelled dataset for training our model.

\subsection{{\DatasetName} Dataset}
\label{sec:datasetv1}
As opposed to {\DatasetNameVZero} which focuses on global conservation sites, {\DatasetName} specifically focuses on conservation sites in India and Nepal. The {\DatasetName} corpus contains 4,137 articles (150 for Nepal and 3,987 for India) from 1,074 conservation sites across the two countries.
All articles were taken in the two-year span from November 2019 to November 2021. We use \texttt{NewsAPI} to search for the official names of the conservation sites, or alternative and/or local names for the sites as recorded at WWF.

Given the data availability as well as the annotator capacity of the local domain experts from India and Nepal, we labeled all 150 articles from Nepal and only 1,000 articles from India. 
Annotation for {\DatasetName} was done along two dimensions: \textit{Conservation Relevance} and \textit{Infrastructure Relevance}. \textit{Conservation Relevance} is similar to the one described for {\DatasetNameVZero} in Section \ref{sec:datasetv0}.
Among the articles which were labelled as positive for \textit{Conservation Relevance}, we further categorize whether it is relevant to infrastructure. This covers issues such as new roads in forested areas and construction projects near national parks. 
Each article was annotated by two domain experts, one from WWF UK, and another from either WWF India or WWF Nepal. We provided the annotators with a descriptive rubric for labeling in each dimension, as well as concrete examples of edge cases. The following was one such example in our instructions:
\begin{formal}
Articles describing tourism or wildlife or natural beauty of a national park, but without talking about environmental impacts or threats to wildlife and conservation, do not count as positive for \textit{Conservation Relevance}.
\end{formal}
Where the two sets of labels disagree, the authors closely inspect the articles and decide on the final labels. 
\section{Relevance Classification Module}
\label{sec:model}
We highlight the structure of our {\ModelName} classification module and other key techniques used during training.

\subsection{Classification Model}
\label{sec:classification}
The backbone of the {\ModelName} classification model is a BERT model \cite{devlin-etal-2019-bert} with a linear classification head.
BERT is a Transformer-based language model trained using masked language modelling and next sentence prediction objectives on large-scale corpora of books and articles. This large-scale pretraining, as well as its ability to effectively encode context, leads to superior performance on a wide variety of tasks. We adapt BERT to the domain of conservation and infrastructure, and we fine-tune it to perform news article classification. In Section \ref{sec:experiments}, we explore different variants of the BERT model (such as RoBERTa).

One key change we make to the BERT model is that in the final linear head after the main BERT layers, instead of only considering the BERT vector outputs, we also incorporate other features, namely sentiment analysis and topic modelling, as shown in Figure \ref{fig:model-fig}. We hypothesize that including these additional features will provide the model with more useful information that will help classify whether or not a particular article is relevant to infrastructure or conservation. For instance, if an article has topic vectors that align with other articles covering forest habitats, but it has an overwhelmingly positive sentiment, then we may suspect that it could be a tourism-related feature article instead of a conservation-related news article (which are often more neutral or negative in terms of sentiment).

For sentiment analysis, we extract the sentence polarity scores of the article title, its description, and its content, giving us three sentiment scores per article. This is done on a scale of $-1.0$ to $+1.0$, with $-1.0$ representing the most negative score and $+1.0$ representing the most positive score. Sentiment analysis was done using the \texttt{textblob} package \cite{loria2018textblob}.
Meanwhile, for topic extraction, we consider the entire training corpora of {\DatasetNameVZero} and {\DatasetName}, and train a Latent Dirichlet Allocation (LDA) model to identify topic clusters. We use 50 topics for the LDA model and implemented it using \texttt{scikit-learn} \cite{scikit-learn}.
Lastly, for the main BERT model, we concatenate the title, description, and content of each article, and we use this concatenated text as input to our classifier. For cases where the article is missing certain features (e.g. no description), we simply supply an empty string for that feature. The vectors from the three steps (i.e. BERT model, sentiment analysis, topic modelling) are then concatenated, and this final vector is used as the input to the final classification head to generate a binary prediction. Specific implementation settings and other hyperparameters can be found in Section \ref{subsec:experiment-settings}.

\subsection{Active Learning}
\label{sec:active-learning}

Annotating a dataset is costly. In curating our {\DatasetName} dataset, we need to be mindful of which specific articles to label in order for our model to learn most efficiently. For this selection process, we first fine-tune a pretrained RoBERTa-base model on the existing {\DatasetNameVZero} dataset, based on the \textit{Classification Relevance}. To make this preliminary model as close to our final model as possible, we also incorporate the topic modelling and sentiment analysis features, as shown in Figure \ref{fig:model-fig}. Because this is only a preliminary model, we forego doing extensive hyperparameter tuning and decided to just select a setting that worked decently well: with a learning rate of 1e-5, batch size of 16, and training for 10 epochs, we were able to get an F-score of 0.61 on {\DatasetNameVZero}. Using this trained model, we then generate \textit{Classification Relevance} predictions for all articles in the {\DatasetName} corpus, together with the corresponding softmax scores. We treat these softmax scores as a measure for the classification confidence of the model: if the softmax is close to 0 or close to 1, then it means that the model is very certain with its prediction, while if the softmax is close to 0.5, then it means the model is unsure with its prediction. 

We then select 300 articles which our model is least confident about. We hypothesize that selecting these ``difficult'' rows will have the greatest impact on model performance. We call this active learning-based dataset {\DatasetNameVOne}. To verify the effectiveness of active learning, we also randomly sample 300 articles to label, which we call {\DatasetNameVTwo}.
We will later evaluate how this compares with the actively selected dataset on a randomly selected test set of 400 samples in our ablation study (Section \ref{subsec:ablation-study}). 

\subsection{Noisy Label Correction}
\label{subsec:noisy-label-correction}
Our dataset is labelled by two sets of domain expert annotators from WWF. Although we provided detailed criteria for labelling each article, there is always room for some subjectivity in the process. This resulted in the two sets of labels not agreeing with each other on over $10\%$ of the data points.
Although, as mentioned in Section~\ref{sec:datasetv1}, we did manage to obtain the ``ground truth'' label for a small subset of {\DatasetName} for model evaluation purposes, doing that for every single article is prohibitively expensive -- much more expensive than the (not cheap) process of having either annotator providing a (noisy) label. Therefore, in order for {\ModelName} to work well once deployed, we need to be able to learn well from the potentially noisy labels only.

More formally, let $x_n$ be the embedding of an article along with its sentiment and topic modeling vectors as described in Section~\ref{sec:classification}. Let $y_n$ be the true label of this article. The task is to make an accurate prediction on the dataset $\{(x_n, y_n): n=1\dots N\}$ when we only have access to the noisy data $\{(x_n, \tilde y_n): n=1\dots N\}$ where $\tilde y_n$ is the label that we get from either of the two annotators, and the true labels $y_n$ are the final labels that we decide on after resolving conflicts.

To address this challenge, we adapt the CORES$^2$ loss \cite{DBLP:conf/iclr/ChengZLGSL21} noise correction algorithm, which is an extension of the earlier peer loss \cite{peer-loss}. 
Peer loss frames the task of learning from noisy labels as a peer prediction problem.
In practice, the loss for each $(x_n, y_n)$ data point can be calculated using the standard cross entropy loss with $(x_n, y_n)$, modified with a loss calculated using a randomly sampled input $x_{n_1}$ and an \textit{independently} randomly sampled label $y_{n_2}$. That is, we have $$\ell_{\tiny{\textnormal{PEER}}}(f(x_n), \tilde{y}_n) := \ell(f(x_n), \tilde{y}_n) - \alpha \cdot \ell(f(x_{n_1}), \tilde{y}_{n_2})$$ where $\alpha > 0$ is a tunable parameter.
Meanwhile, CORES$^2$ replaces the random sampling from peer loss with a confidence regularizer defined as follows: 
$$\ell_{\tiny{\textnormal{CORES}}}(f(x_n), \tilde{y}_n) := \ell(f(x_n), \tilde{y}_n) - {\beta} \cdot \mathbb{E}_{\mathcal{D}_{\tilde{Y}|\tilde{D}}} [\ell (f(x_n), \tilde{Y})]$$ where $\tilde{D}$ is the dataset, $\tilde{Y}$ is a noisy label, and $\beta > 0$ is a tunable parameter. Following \citet{DBLP:conf/iclr/ChengZLGSL21}, we calculate this confidence regularizer term using an estimate of the noise prior probability.
We test both peer loss and CORES$^2$ loss, and report results in our ablation study (Section \ref{subsec:ablation-study}).
\section{Article Postprocessing Module}
\label{sec:article-postprocessing-module}
Once the relevant articles are identified using the model, we then perform a few post-processing steps to extract key information and make them easier to analyze and visualize.

\subsection{Keyword Extractor}
\label{subsec:keyword-extractor}
Keywords are important, as they allow the easy summarization, categorization, and grouping of news articles. Furthermore, we also use these keywords as hashtags in our social media module (Section \ref{sec:deployment}). To extract keywords, we use an extensive list of conservation-related keywords maintained at WWF and search the article for exact matches. In addition, we also use Named Entity Recognition systems to extract the salient words in each article. To perform this, we use a BERT-based model trained on the CoNLL 2003 Named Entity Recognition dataset \cite{tjong-kim-sang-de-meulder-2003-introduction}. The keywords extracted using these two methods are then concatenated to form the final set of keywords.

\subsection{Event Extractor}
\label{subsec:event-extractor}
\begin{figure}
    \centering
    \includegraphics[width=\columnwidth]{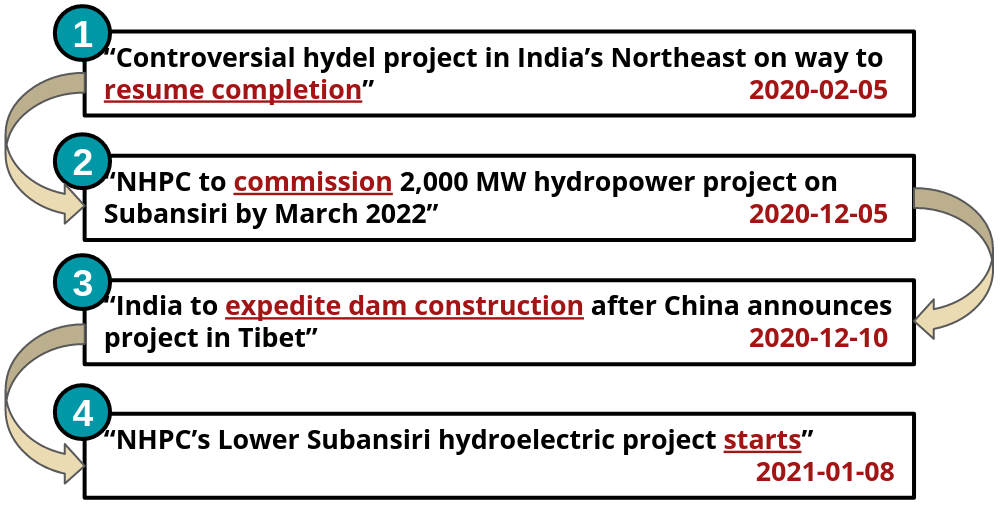}
    \caption{Example of events selected by the Event Extractor (Section \ref{subsec:event-extractor}) by date. The progression of the project is highlighted by the phrases in red underline.}
    \label{fig:event-fig}
\end{figure}

To track the progress of infrastructure projects, it is often not enough to just view a single article in isolation. Rather, news regarding these projects often builds up over a period of weeks or months. To help provide this context, we create an automated event extractor, which leverages our {\DatasetName} dataset, including both the labelled articles as well as the unlabelled articles. Given a new article $a$, our goal is to find past articles $P_a$ which are closely related to $a$. We first gather all previous articles which are from the same conservation site. Next, we create a graph $G_a$, where each article is a node, and two nodes share an edge if the corresponding articles share $\geq k$ common keywords (from Section \ref{subsec:keyword-extractor}). Here, $k$ is an adjustable parameter depending on how loosely connected we want $G_a$ to be. For our data, we use $k=3$. Once the graph $G_a$ is constructed, we then define an ``event'' to be the maximal clique containing $a$, and we report all such events. A sample chain of events is shown in Figure \ref{fig:event-fig}.

\subsection{Geolocation}
\label{subsec:geolocation}
To aid with visualization (Section \ref{sec:deployment}), we perform geolocation on the classified news articles, based on the search terms used to retrieve them.
To extract latitude and longitude coordinates, we leverage an extensive directory of conservation sites from WWF, and we use the directory to map conservation sites to their corresponding coordinates. If the directory contains no match, we geolocate using the \texttt{geopy} package.

\section{Experiments and Results}
\label{sec:experiments}
Here, we discuss results of our in-lab experiments and ablation studies to verify our hypotheses. Results from real-world deployment are discussed in the succeeding section.

\subsection{Experiment Settings}
\label{subsec:experiment-settings}
\subsubsection{Baselines}
We compare the performance of our {\ModelName} model with the following baselines:
\begin{enumerate}
    \item \textbf{Keyword model}: We consider a naive model that checks for the count of certain keywords. We curate two sets of ``conservation-related keywords'' and ``infrastructure-related keywords''. If an article contains more than $k$ ``conservation-related keywords'', then it is considered to be relevant to conservation (likewise for infrastructure).
    \item \textbf{RNN-based models}: We tokenize each article, then pass the embedding to RNN models, where the hidden state of the last layer is used as input to the final classification layer. We use two types of RNN models, namely GRUs \cite{bahdanau2014neural} and LSTMs \cite{HochSchm97}. 
    \item \textbf{BERT-based models}: We fine-tune a pretrained BERT-base \cite{devlin-etal-2019-bert} and RoBERTa-base model \cite{Liu2019RoBERTaAR}, where we add a classification head after the final layer to perform relevance classification.
\end{enumerate}

\subsubsection{Evaluation Metrics}
Since our task is binary classification, we measure the accuracy, precision, recall, and F1-score. For precision, recall, and F1, we consider only the scores of the positive class. All metrics are calculated separately for \textit{Conservation Relevance} and \textit{Infrastructure Relevance}.

\subsubsection{Data}
For \textit{Conservation Relevance}, we train on the {\DatasetName} dataset (consisting of both {\DatasetNameVOne} and {\DatasetNameVTwo}), as well as the {\DatasetNameVZero} dataset. For \textit{Infrastructure Relevance}, since {\DatasetNameVZero} does not contain infrastructure labels, we only train using {\DatasetName}. We split the training data into an 80-20 training-validation split. For evaluation, we use the test split of {\DatasetName} for both \textit{Conservation Relevance} and \textit{Infrastructure Relevance}.

\subsubsection{Implementation Settings}
\label{sec:implementation-settings}
For the GRU/LSTM, we use a batch size of 128, hidden size of 128, and dropout of 0.2. We train for 10 epochs with a learning rate of 1e-4. Meanwhile, for BERT, RoBERTa, and {\ModelName}, we train for 10 epochs with batch size 4 and learning rate 1e-5. We use RoBERTa for the backbone model of {\ModelName}. 
Model selection is done by considering the best validation F1-score. 

\subsection{Results and Analysis}
\label{subsec:results-and-analysis}

\begin{table}[t]
	\begin{subtable}[t]{\columnwidth}
		\centering
		\begin{tabular}{c|c|c|c|c}
            \hline 
            \textbf{Model} & \textbf{Acc.} & \textbf{P} & \textbf{R} & \textbf{F1} \\
            \hline \hline
            Keyword & 0.820 & 0.317 & 0.634 & 0.423 \\
            LSTM & 0.711 & 0.495 & 0.511 & 0.504 \\
            GRU & 0.729 & 0.422 & 0.505 & 0.475 \\
            BERT & 0.860 & 0.708 & 0.704 & 0.706 \\
            RoBERTa & 0.867 & 0.705 & 0.743 & 0.721 \\
            \ModelName & \textbf{0.877} & \textbf{0.729} & \textbf{0.801} & \textbf{0.744} \\
            \hline
        \end{tabular}
		\caption{Scores for \textit{Conservation Relevance}}
		\label{tab:conservation-results}
	\end{subtable}
	\hfill
	\begin{subtable}[t]{\columnwidth}
		\centering
		\begin{tabular}{c|c|c|c|c}
        \hline 
            \textbf{Model} & \textbf{Acc.} & \textbf{P} & \textbf{R} & \textbf{F1} \\
            \hline \hline
            Keyword & \textbf{0.947} & 0.250 & 0.455 & 0.323 \\
            LSTM & 0.908 & 0.566 & 0.537 & 0.554 \\
            GRU & 0.895 & 0.544 & 0.557 & 0.553 \\
            BERT & 0.922 & 0.840 & 0.745 & 0.771 \\
            RoBERTa & 0.916 & 0.794 & 0.809 & 0.799 \\
            {\ModelName} & 0.941 & \textbf{0.880} & \textbf{0.821} & \textbf{0.850} \\
            \hline
        \end{tabular}
		\caption{Scores for \textit{Infrastructure Relevance}}
		\label{tab:infrastructure-results}
	\end{subtable}
	\caption{Average scores for \textit{Conservation Relevance} (Table \ref{tab:conservation-results}) and \textit{Infrastructure Relevance} (Table \ref{tab:infrastructure-results}), taken over 10 random seeds.}
	\label{tab:main-results-table}
\end{table}

Experimental results are shown in Tables \ref{tab:conservation-results} and \ref{tab:infrastructure-results}. We observe that indeed, adding the sentiment analysis and topic modelling features, as well as the CORES$^2$ loss for noisy label correction, aids in predictions for both \textit{Conservation Relevance} and \textit{Infrastructure Relevance}, providing an improvement over both BERT-base and RoBERTa-base. 

Our data is quite imbalanced: $>$80\% of the articles are not relevant. This manifests itself in the discrepancies between accuracy and F1-score. We observe, for example, that the naive keyword model has very high accuracy scores but very low F1-scores, which indicates that it predicts a lot of zeros (hence the high accuracy), but is not able to predict the relevant articles well. The RNN-based models (LSTM and GRU) seem to perform relatively poorly, achieving an F1-score of around 0.5. This could also be attributed to the data imbalance, since these RNN-based models are generally not as robust to imbalanced datasets. In contrast, the BERT and RoBERTa models perform quite well, with F1-scores $>$0.7 for conservation and $>$0.75 for infrastructure, and precision/recall scores also around that range. This indicates that these transformer-based models are able to generalize quite well and successfully capture the notions of \textit{Conservation Relevance} and \textit{Infrastructure Relevance}. Lastly, {\ModelName} offers significant improvement over the RoBERTa-base model (F1 t-test $p$-value $=0.018$ for conservation and $0.033$ for infrastructure), showing the positive effects of incorporating information such as the emotion and topics over simply considering the article text in isolation.

\subsection{Ablation Study}
\label{subsec:ablation-study}

\subsubsection{Active Learning} 
We compare the effect with training on actively-sampled data ({\DatasetNameVOne}) and randomly-sampled data ({\DatasetNameVTwo}). 
Each of these datasets contain 300 India articles, as detailed in Section~\ref{sec:active-learning} and~\ref{sec:datasetv1}.
We append these articles to the existing {\DatasetNameVZero} to create the final data for training. We use the RoBERTa model for these experiments. Results are shown in Table \ref{tab:ablation-active-learning}.

For both {\DatasetNameVOne} and {\DatasetNameVTwo}, we see an improvement over just using {\DatasetNameVZero}. Indeed, training with more data will result in better performance, regardless of how the data is sampled. We also observe that adding actively sampled data results in a larger improvement than adding randomly sampled data across all metrics (F1 t-test $p$-value $=0.004$). This verifies the effectiveness of our hypothesized confidence-based data selection for annotation.

\begin{table}[t]
    \centering
    \begin{tabular}{c|c|c|c|c}
        \hline 
        \textbf{Dataset} & \textbf{Acc.} & \textbf{P} & \textbf{R} & \textbf{F1} \\
        \hline \hline
        {\DatasetNameVZero} & 0.911 & 0.585 & 0.585 & 0.586 \\
        \hline
        \textsc{WHS+Inf.Corp-A} & \textbf{0.921} & \textbf{0.600} & \textbf{0.774} & \textbf{0.670} \\
        \textsc{WHS+Inf.Corp-R} & 0.916 & 0.586 & 0.696 & 0.637 \\
        \hline
    \end{tabular}
    \caption{Evaluation scores for \textit{Conservation Relevance} for {\DatasetNameVOne} compared with {\DatasetNameVTwo}, averaged over 10 random seeds.}
    \label{tab:ablation-active-learning}
\end{table}

\begin{table}[t]
    \centering
    \begin{tabular}{c|c|c|c|c}
        \hline 
        \begin{tabular}[t]{@{}c@{}} \textbf{Noisy Label} \\ \textbf{Correction} \end{tabular} & \textbf{Acc.} & \textbf{P} & \textbf{R} & \textbf{F1} \\
        \hline \hline
        None & 0.907 & 0.566 & 0.441 & 0.497 \\
        \hline
        Peer Loss & \textbf{0.911} & \textbf{0.591} & 0.465 & 0.509 \\
        CORES$^2$ & 0.908 & 0.584 & \textbf{0.551} & \textbf{0.553} \\
        \hline
    \end{tabular}
    \caption{Evaluation scores for \textit{Conservation Relevance} for two noise correction methods, over 10 random seeds.}
    \label{tab:ablation-noisy-label}
\end{table}

\subsubsection{Noisy Label Correction} We examine the effect of the noise correction methods outlined in Section \ref{subsec:noisy-label-correction}, by comparing the effect of using peer loss, CORES$^2$ loss, and standard cross entropy loss. Based on {\DatasetName}, we use the labels supplied by one of the two annotators for the training set, and the final calibrated labels for the test set. Hyperparameter search was done for both peer loss and CORES$^2$ loss to find the optimal values of $\alpha=0.05$ and $\beta=0.05$. We trained for 20 epochs with a learning rate of 2e-5.

From Table \ref{tab:ablation-noisy-label}, we observe that for accuracy and precision, all three losses perform very similarly, with peer loss performing the highest by a small margin. For recall and F1, peer loss and the standard loss perform at comparable levels, while CORES$^2$ loss performs better than both (F1 t-test $p$-value $=0.001$). This is likely because the confidence regularizer used in CORES$^2$ works better than the random sampling used by peer loss. 
Both peer and CORES$^2$ loss might work even better if we had more training data than the current 600 in {\DatasetName}. In the end, given the positive results of CORES$^2$, we used it in our {\ModelName} model.
\section{Deployment and Impact}
\label{sec:deployment}

\ModelName{} has been used at WWF since February 2022. We describe the deployment, results, and lessons learned.

\subsection{Pilot Study}
\label{subsec:pilot-study}

The first stage of {\ModelName} deployment, which is the pilot study, started in February 2022 and ran for around one month. Every week, the CMU team scraped the news articles and ran the entire {\ModelName} pipeline, forwarding the outputs to the WWF teams to examine and provide feedback. During this pilot phase, the WWF and CMU teams identified a range of operational and technical issues in the initial version of \ModelName{}.

\begin{figure*}
    \centering
    \includegraphics[width=0.24\textwidth]{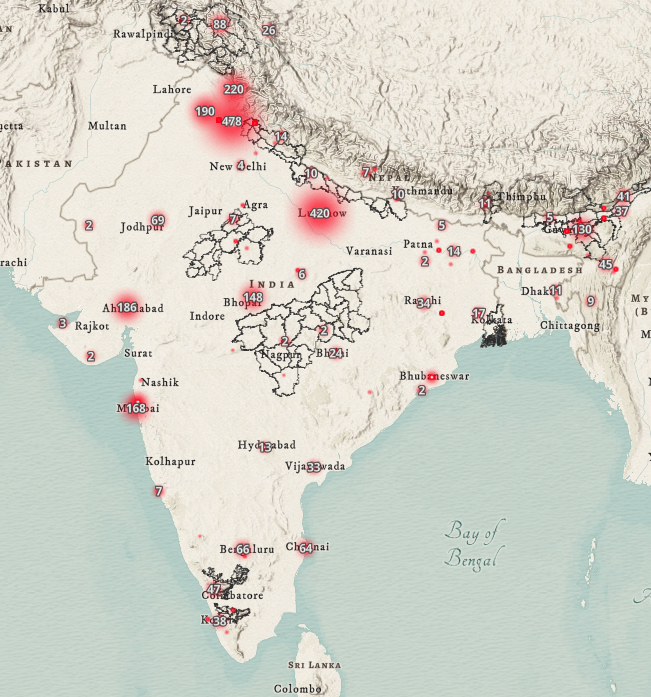}
    \includegraphics[width=0.75\textwidth]{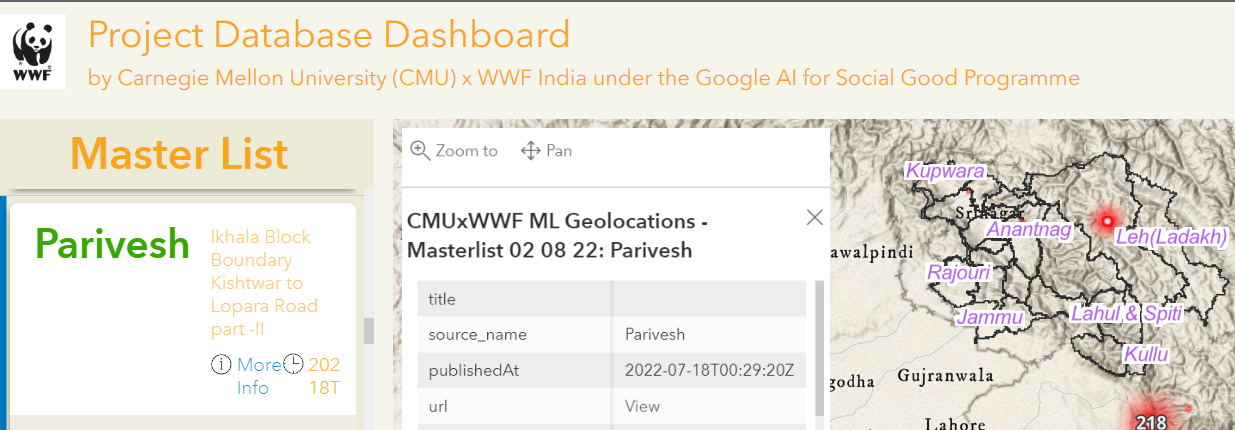}
    \caption{Left: The highlighted red areas indicate clusters of articles found by our model. Right: The WWF GIS system, where each relevant article is shown on the map with its corresponding key details.}
    \label{fig:gis}
\end{figure*}

First, in order for \ModelName{} to fit into the established workflow of WWF, it needs to be integrated into its GIS system. During the pilot, we realized that it is crucial to add the geolocation of each article (Section \ref{subsec:geolocation}) and format the model output according to the specifications of the GIS platform used at WWF. Figure~\ref{fig:gis} shows how \ModelName{}'s results get integrated into the GIS system, with the red areas being the locations where we identify a relevant article.

We also discovered that while \texttt{NewsAPI} has a good collection of global news sources, it fails to include some relevant sources in the local context. With the suggestions from the WWF team, we incorporated additional sources 
that often yield relevant local articles. One such site is Parivesh, which contains proposals of infrastructure projects in India.

Finally, we found that some conservation sites' names often lead to 0 results, while other terms were too general and yielded hundreds of results, almost all of which were irrelevant, leading to inefficiencies. We set a lower and upper threshold, and filter out search terms outside the thresholds.

\subsection{Deployment Results}
After we resolved the above issues, we proceeded with the actual deployment. The procedure was similar to the pilot phase, except that at this phase, the focus is to evaluate the performance of \ModelName{}. The WWF teams closely inspected the model predictions each week and provided ground truth labels for each article. The label feedback allowed the CMU team to retrain the model regularly. This stage ran from March 2022 to July 2022. Table \ref{tab:deployment-results} shows the aggregated results over 5 months of evaluation results from WWF India, Nepal, and UK. WWF UK labeled the first half of the deployment for all locations and India/Nepal labeled the second half for news articles in their respective countries.

Overall, {\ModelName} continued to show great performance in \textit{Conservation Relevance} during real-world deployment. 
Across all evaluations, the precision scores are consistently high, indicating that almost all of the articles reported by {\ModelName} are indeed relevant. We intentionally tuned the model towards this direction -- when almost everything that the model flagged is relevant, it would greatly help with establishing the trust in the model at the early stage of deployment. 
As we continue developing the model, we aim to improve the model towards achieving higher recall, to be able to capture more relevant articles.

On the other hand, on \textit{Infrastructure Relevance} for India, the model's performance was worse than the offline experiments. Upon further inspection, we discovered that the majority of mistakes were in fact only 2-4 original pieces of news that were paraphrased by various news sources into 20-40 articles. Since there are only a few \textit{Infrastructure Relevance} positive articles to start with, this had a big impact on the model performance. Meanwhile, such phenomenon did not occur in our offline experiments because there we randomly sampled news from a large corpus for labeling.

Aside from overall metrics, we also highlight individual success stories.
Figure \ref{fig:gis}(right) shows a concrete example where {\ModelName} made a difference. 
In early August, 2022, {\ModelName} detected a new project of Ikhala Block Boundary Kishtwar to Lopara Road and highlighted it in the WWF GIS system. Upon further investigation by WWF staff, it is found that the project would divert 5.9 hectares of forest land. More importantly, WWF found that the project was still at its pre-proposal stage. This means WWF would be able to take early action and possibly participate in relevant conversations.
Such stories are happening frequently since the deployment of {\ModelName}.
Using the tool's outputs integrated into our internal GIS systems, the WWF staff are continuously coordinating with our field teams to examine the status and report on relevant projects and areas.

\subsection{Qualitative and Quantitative Comparison with Current Practice}

Prior to \ModelName{}, WWF had already been monitoring media for conservation-related articles (Figure~\ref{fig:page-1-fig}). However, most of these efforts were not very structured or logged. It is thus difficult to draw head-to-head comparisons between \ModelName{} and WWF's existing approach. That said, we still provide qualitative and quantitative evidence supporting the merit of \ModelName{} over the current practice.

\begin{table}[t]
    \small
    \centering
    \begin{tabular}{c|ccc|ccc}
        \hline
         & \multicolumn{3}{c}{\textbf{{Conservation}}} & \multicolumn{3}{c}{\textbf{{Infrastructure}}} \\
         & P & R & F1 & P & R & F1 \\
        \hline \hline
        India & 0.849 & 0.605 & 0.706 & 0.462 & 0.250 & 0.324 \\
        Nepal & 0.895 & 0.917 & 0.906 & 0.923 & 0.308 & 0.462 \\
        UK & 0.879 & 0.823 & 0.850 & 1.000 & 0.455 & 0.625 \\   
        \hline
    \end{tabular}
    \caption{Aggregated scores of \textsc{NewsPanda} on weekly articles from March 2022 to July 2022.}
    \label{tab:deployment-results}
\end{table}

Two months into the deployment, the CMU team carried out semi-structured interviews with their WWF colleagues who have been using \ModelName{} outputs in their work. The purpose was to understand how WWF teams liked the toolkit and to elicit possible suggestions for improvement. Some quotes from the interviews are as follows.

\begin{formal}
``You're giving us a bunch of articles... over 50 articles a week. We had two interns who spend 2-3 days a week on this and would only give us seven to ten articles. So there is a huge bump in efficiency right there in itself.''
\end{formal}

\begin{formal}
``The data that you're sharing give a global perspective. It is very useful to understand the upcoming projects or mitigation measures that are being adopted on a global scale. So it helps us be informed.''
\end{formal}

This improvement in news collection also helped with the downstream task -- infrastructure impact assessment.

\begin{formal}
``It took us maybe a month to do analyses of three or four infrastructure projects. With \ModelName{}, we can send (stakeholders) 20 or 30 reports in a month.''
\end{formal}

The micro-level improvement in this single task has also resulted in macro-level organizational change:

\begin{formal}
``It's also a transition in their (WWF staff) job function. They will not just be doing data hunting. They are qualifying themselves to be data analysts.''
\end{formal}

The WWF Nepal team has been putting together weekly news digests for conservation sites in Nepal. Although this dataset is small and has no negative labels, this is the only quantitative comparison between \ModelName{} and current practice we can make. We find that our model is able to identify 62\% of the articles in the news digest. This is a relatively good performance as we had extremely limited articles (only 150) about Nepali conservation sites to train the model.

\subsection{Sustainable Deployment and Broader Impact}
Encouraged by the success of \ModelName{} at the initial stages, we are working to scale it to more sites and permanently deploy \ModelName{} as part of the WWF computing infrastructure. We have been collecting news articles for over 60,000 sites globally and applying our trained model to classify them on a weekly basis since April 2022.
Because the main model has already been trained, we no longer need extensive data labeling for evaluation. Instead, we only need a small subset for model update and fine-tuning purposes. We are currently investigating the ability \ModelName{} to generalize to new locations and new languages given only a few (or even zero) domain-specific training points.
We are also shifting our system to a cloud server to be owned and maintained by the WWF team, rather than the CMU team, to ensure sustainable deployment. The CMU team will continue to provide support and tutorials to help WWF eventually grow in-house capability of sustaining the project. 

\begin{figure}
    \centering
    \includegraphics[width=\columnwidth]{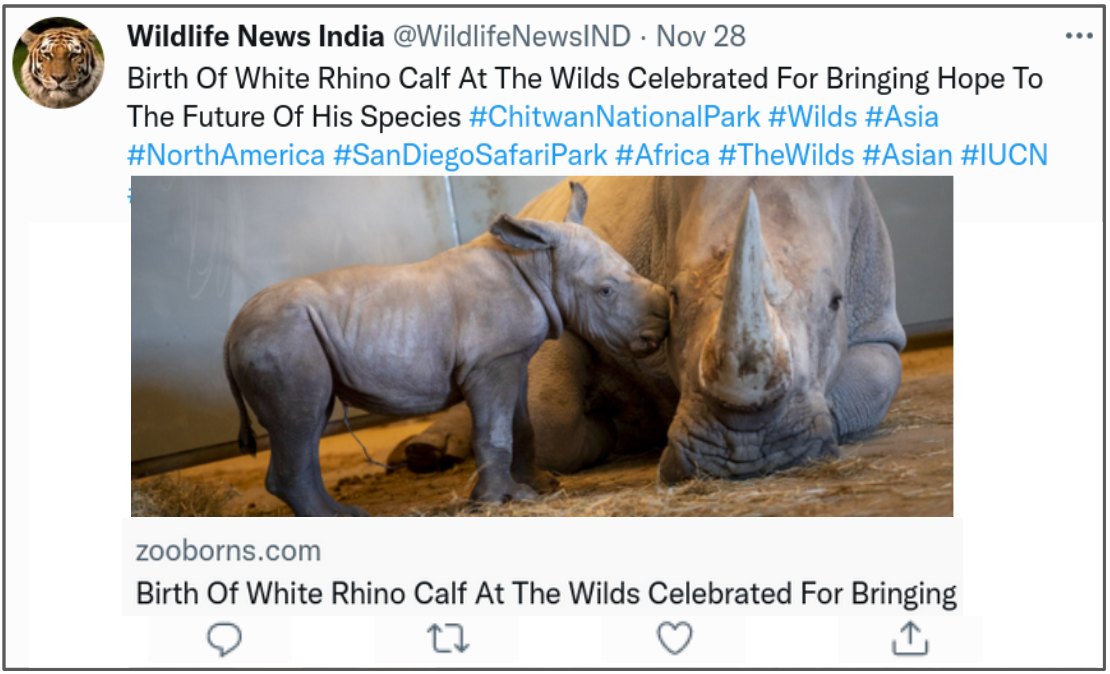}
    \caption{Sample tweet of {\BotName}}
    \label{fig:twitter}
\end{figure}

Much as this project was a collaboration between WWF and CMU, \ModelName{} could also be valuable to the broader civil society. Thus, we also developed a social media module in the form of a Twitter bot called {\BotName}. The bot periodically tweets a selected set of the identified relevant articles. In addition to tweeting links to articles, we also use the keywords from {\ModelName}'s keyword extractor (Section \ref{subsec:keyword-extractor}) to generate salient hashtags. A sample tweet is shown in Figure \ref{fig:twitter}. Currently, {\BotName} is focused on conservation-related articles in India. As we continue working on this project, we hope to scale this to a global level, so that any organization or individual interested in conservation can benefit from the tool.

\subsection{Lessons Learned}
This 1.5 year long and counting collaboration has yielded many valuable lessons for both WWF and CMU. We have already mentioned some of those in earlier sections.
We highlight two more generalizable lessons below.

Problem identification is an iterative process and rapid prototyping helps surface unforeseen needs. The event extractor in Section~\ref{subsec:event-extractor} was not initially part of the agenda: without a prototype of the model readily available, it was difficult for WWF to realize what could be done with it. However, after several iterations of communication and exploring results, the need to track the development related to a single project/location became clear to us. This was made possible by the rapid prototyping, where the CMU team used viable algorithms that may not be optimal but are quick to implement to demonstrate the possibilities of the toolkit.

It is the various ``not-so-AI'' components that realize the promise of an AI for nonprofit project on the ground. While the classification module in Section~\ref{sec:model} is the engine of {\ModelName}, the postprocessing module in Section~\ref{sec:article-postprocessing-module} and the visualization module in Figure~\ref{fig:gis} are key in getting the information in a consumable format, and ultimately the buy-in at WWF. Each of the latter two modules requires at least as much engineering effort and careful design as the classification module. We call on future AI for nonprofit projects to pay enough attention to all the infrastructure around the AI part, in order to deliver the real impact that we hoped for.
\section{Conclusion}
In this paper, we designed and deployed {\ModelName}, a toolkit for extracting, classifying, and analyzing articles related to conservation and infrastructure. We showed empirically that our {\ModelName} model classifies better than baseline methods for both \textit{Conservation Relevance} and \textit{Infrastructure Relevance}. We also presented quantitative and qualitative evaluations of our system in the real world as well as its impact on WWF teams in UK, India, and Nepal. 

Currently {\ModelName} mainly focuses on a few countries, and we are expanding it to a global scale. 
However, incorporating additional conservation sites is just the beginning. 
To do it right,
we also need to cover more languages and more local media sources. This is especially important for the global south, as many high-impact local developments might never reach international news outlets.
The ability to capture these local sources, especially if they are not written in English, is something we are currently working on. 
We are currently starting with articles written in the Nepali language. Initial experiments with a multilingual version of {\ModelName} have shown good generalization when given only a few Nepali articles for training. With this multilingual model, we hope to further expand to cover a wider array of languages. 

\section*{Acknowledgements}
We thank Mr. Pramod Neupane, Consultant-Sustainable Infrastructure at World Bank, for all his support during the initial phase of the project which includes project conceptualization, data curation, funding acquisition, project administration for WWF Nepal, and resources allocation at WWF Nepal. We also thank the communications team at WWF Nepal for providing the weekly news links.  This work was supported in part by a Google AI for Social Good award, NSF grant IIS-2046640, a Siebel Scholarship and a Carnegie Mellon Presidential Fellowship.

\bibliography{aaai23}

\begin{thebibliography}{19}
\providecommand{\natexlab}[1]{#1}

\bibitem[{Bahdanau, Cho, and Bengio(2014)}]{bahdanau2014neural}
Bahdanau, D.; Cho, K.; and Bengio, Y. 2014.
\newblock Neural machine translation by jointly learning to align and
  translate.
\newblock \emph{arXiv preprint arXiv:1409.0473}.

\bibitem[{Boutilier and Bahr(2020)}]{Boutilier2020}
Boutilier, R.; and Bahr, K. 2020.
\newblock A Natural Language Processing Approach to Social License Management.
\newblock \emph{Sustainability}, 12.

\bibitem[{Cheng et~al.(2021)Cheng, Zhu, Li, Gong, Sun, and
  Liu}]{DBLP:conf/iclr/ChengZLGSL21}
Cheng, H.; Zhu, Z.; Li, X.; Gong, Y.; Sun, X.; and Liu, Y. 2021.
\newblock Learning with Instance-Dependent Label Noise: A Sample Sieve
  Approach.
\newblock In \emph{ICLR}.

\bibitem[{Devlin et~al.(2019)Devlin, Chang, Lee, and
  Toutanova}]{devlin-etal-2019-bert}
Devlin, J.; Chang, M.-W.; Lee, K.; and Toutanova, K. 2019.
\newblock {BERT}: Pre-training of Deep Bidirectional Transformers for Language
  Understanding.
\newblock In \emph{Proceedings of the 2019 Conference of the North {A}merican
  Chapter of the Association for Computational Linguistics: Human Language
  Technologies, Volume 1 (Long and Short Papers)}, 4171--4186. Minneapolis,
  Minnesota: Association for Computational Linguistics.

\bibitem[{Hochreiter and Schmidhuber(1997)}]{HochSchm97}
Hochreiter, S.; and Schmidhuber, J. 1997.
\newblock Long Short-Term Memory.
\newblock \emph{Neural Computation}, 9(8): 1735--1780.

\bibitem[{Hosseini and Coll~Ardanuy(2020)}]{data_study_group_team_2020_3878457}
Hosseini, K.; and Coll~Ardanuy, M. 2020.
\newblock Data Study Group Final Report: WWF.

\bibitem[{Joshi, N, and Rao(2016)}]{Joshi2016}
Joshi, K.; N, B.; and Rao, J. 2016.
\newblock Stock Trend Prediction Using News Sentiment Analysis.
\newblock \emph{International Journal of Computer Science and Information
  Technology}, 8: 67--76.

\bibitem[{Lisivick(2018)}]{newsapi}
Lisivick, M. 2018.
\newblock NewsAPI Python Library.
\newblock \url{https://github.com/mattlisiv/newsapi-python}.
\newblock Accessed: 2022-12-12.

\bibitem[{Liu and Guo(2020)}]{peer-loss}
Liu, Y.; and Guo, H. 2020.
\newblock Peer Loss Functions: Learning from Noisy Labels without Knowing Noise
  Rates.
\newblock In \emph{Proceedings of the 37th International Conference on Machine
  Learning}, ICML'20. JMLR.org.

\bibitem[{Liu et~al.(2019)Liu, Ott, Goyal, Du, Joshi, Chen, Levy, Lewis,
  Zettlemoyer, and Stoyanov}]{Liu2019RoBERTaAR}
Liu, Y.; Ott, M.; Goyal, N.; Du, J.; Joshi, M.; Chen, D.; Levy, O.; Lewis, M.;
  Zettlemoyer, L.; and Stoyanov, V. 2019.
\newblock RoBERTa: A Robustly Optimized BERT Pretraining Approach.
\newblock \emph{ArXiv}, abs/1907.11692.

\bibitem[{Loria(2018)}]{loria2018textblob}
Loria, S. 2018.
\newblock textblob Documentation.
\newblock \emph{Release 0.15}, 2.

\bibitem[{Luccioni, Baylor, and Duchene(2020)}]{luccioni2020analyzing}
Luccioni, S.; Baylor, E.; and Duchene, N. 2020.
\newblock Analyzing Sustainability Reports Using Natural Language Processing.
\newblock In \emph{NeurIPS 2020 Workshop on Tackling Climate Change with
  Machine Learning}.

\bibitem[{Murray et~al.(2020)Murray, Gupta, Burke, Rupam, and
  Tshankie}]{omdena}
Murray, L.~C.; Gupta, N.; Burke, J.; Rupam, R.; and Tshankie, Z. 2020.
\newblock Matching Land Conflict Events to Government Policies via Machine
  Learning Models.
\newblock
  \url{https://omdena.com/wp-content/uploads/2019/12/Omdena-Land-Conflicts-Challenge-1.pdf}.
\newblock Accessed: 2022-12-12.

\bibitem[{Ojokoh(2012)}]{Ojokoh2012}
Ojokoh, B. 2012.
\newblock Automated Online News Content Extraction.
\newblock \emph{International Journal of Computer Science Research and
  Application}, 2: 2--12.

\bibitem[{Pedregosa et~al.(2011)Pedregosa, Varoquaux, Gramfort, Michel,
  Thirion, Grisel, Blondel, Prettenhofer, Weiss, Dubourg, Vanderplas, Passos,
  Cournapeau, Brucher, Perrot, and Duchesnay}]{scikit-learn}
Pedregosa, F.; Varoquaux, G.; Gramfort, A.; Michel, V.; Thirion, B.; Grisel,
  O.; Blondel, M.; Prettenhofer, P.; Weiss, R.; Dubourg, V.; Vanderplas, J.;
  Passos, A.; Cournapeau, D.; Brucher, M.; Perrot, M.; and Duchesnay, E. 2011.
\newblock Scikit-learn: Machine Learning in {P}ython.
\newblock \emph{Journal of Machine Learning Research}, 12: 2825--2830.

\bibitem[{Reis et~al.(2004)Reis, Golgher, Silva, and
  Laender}]{10.1145/988672.988740}
Reis, D.~C.; Golgher, P.~B.; Silva, A.~S.; and Laender, A.~F. 2004.
\newblock Automatic Web News Extraction Using Tree Edit Distance.
\newblock In \emph{Proceedings of the 13th International Conference on World
  Wide Web}, WWW '04, 502–511. New York, NY, USA: Association for Computing
  Machinery.
\newblock ISBN 158113844X.

\bibitem[{Santos and Crowder(2021)}]{10.1093/biosci/biaa175}
Santos, B.~S.; and Crowder, L.~B. 2021.
\newblock {Online News Media Coverage of Sea Turtles and Their Conservation}.
\newblock \emph{BioScience}, 71(3): 305--313.

\bibitem[{Tjong Kim~Sang and
  De~Meulder(2003)}]{tjong-kim-sang-de-meulder-2003-introduction}
Tjong Kim~Sang, E.~F.; and De~Meulder, F. 2003.
\newblock Introduction to the {C}o{NLL}-2003 Shared Task: Language-Independent
  Named Entity Recognition.
\newblock In \emph{Proceedings of the Seventh Conference on Natural Language
  Learning at {HLT}-{NAACL} 2003}, 142--147.

\bibitem[{Wu et~al.(2018)Wu, Xie, Huang, Li, Yuan, and Liu}]{wu2018using}
Wu, Y.; Xie, L.; Huang, S.-L.; Li, P.; Yuan, Z.; and Liu, W. 2018.
\newblock Using social media to strengthen public awareness of wildlife
  conservation.
\newblock \emph{Ocean \& Coastal Management}, 153: 76--83.

\end{thebibliography}

\end{document}